\documentclass[12pt,a4paper]{article}
\pdfoutput=1

\usepackage{hyperref}

\usepackage[normalem]{ulem}
\usepackage{cite}
\usepackage{appendix}
\usepackage{epsfig}
\usepackage{amsmath}
\usepackage{amssymb}
\usepackage{bm}
\usepackage{hyperref}
\usepackage{slashed}
\newcommand{\be} {\begin{equation}}
\newcommand{\ee} {\end{equation}}
\newcommand{\ba} {\begin{eqnarray}}
\newcommand{\ea} {\end{eqnarray}}

\newcommand{\cL} {\mathcal L}

\newcommand{\GeV}{\text{GeV}}
\newcommand{\TeV}{\text{TeV}}
\newcommand{\SO}{\text{SO}}
\newcommand{\SU}{\text{SU}}
\newcommand{\U}{\text{U}}

\usepackage{color}
\definecolor{darkblue}{cmyk}{1,0.3,0,0.2}
\definecolor{violet}{cmyk}{0,1,0,0.2}
\hypersetup{colorlinks, bookmarksnumbered, citecolor=darkblue, linkcolor=darkblue, pdfstartview=FitH, urlcolor=darkblue, linktocpage}

\newcommand{\arXhref}[1]{\href{http://arxiv.org/abs/#1}{#1}}

\begin{document}

\begin{flushright}
 ZU-TH-46/15
\end{flushright}

\thispagestyle{empty}

\bigskip

\begin{center}
\vspace{0.5cm}
   {\Large\bf Knocking on New Physics' door \\[0.2cm] with a Scalar Resonance} \\[1cm]
   {\bf Dario Buttazzo$^a$, Admir Greljo$^{a,b}$, David Marzocca$^a$}    \\[0.5cm]
  {\em $(a)$  Physik-Institut, Universit\"at Z\"urich, CH-8057 Z\"urich, Switzerland}  \\ 
  {\em $(b)$  Faculty of Science, University of Sarajevo, Zmaja od Bosne 33-35, 71000 Sarajevo, Bosnia and Herzegovina }   \\[2.0cm]
\end{center}

\centerline{\large\bf Abstract}
\begin{quote}
\indent
We speculate about the origin of the recent excess at $\sim 750$~GeV in diphoton resonance searches observed by the ATLAS and CMS experiments using the first $13$~TeV data. Its interpretation as a new scalar resonance produced in gluon fusion and decaying to photons is consistent with all relevant exclusion bounds from the 8 TeV LHC run. We provide a simple phenomenological framework to parametrize the properties of the new resonance and show in a model-independent way that, if the scalar is produced in gluon fusion, additional new colored and charged particles are required.
Finally, we discuss some interpretations in various concrete setups, such as a singlet (pseudo-) scalar, composite Higgs, and the MSSM.
\end{quote}
\vspace{5mm}

\newpage
\tableofcontents

\section{Introduction}

Very recently, the ATLAS and CMS collaborations presented first results from 13~TeV proton--proton collisions at LHC Run-II \cite{CMStalk,ATLAStalk}. Intriguingly, both experiments found a resonance-like excess in the diphoton invariant mass spectrum around 750 GeV.

The CMS collaboration reported a 95\% CL upper limit of 13.7 fb on the cross section times branching ratio of a narrow spin-2 resonance decaying into two photons, compared with an expected exclusion of 6.3 fb (Fig.~6 of Ref.~\cite{CMStalk}), which corresponds to an excess with a local significance of $2.6\,\sigma$~\cite{CMStalk}.
When interpreting the excess in terms of the signal strength for a narrow scalar (or pseudo-scalar) resonance,
based on the expected and observed exclusion limits, the CMS measurement in the Gaussian approximation reads
\begin{equation}
\mu^{\rm CMS}_{\rm{13 TeV}} = \sigma(pp \to S)_{13\,{\rm TeV}} \times \mathcal{B}(S \to \gamma\gamma)
= (5.6 \pm 2.4) ~\rm{fb}\,.
\label{eq:Excess}
\end{equation}
Moreover, the ATLAS collaboration reported the observed exclusion limit in the fiducial region for a narrow-width scalar resonance $\mu^{\rm ATLAS}_{\rm{13 TeV,fid}} < 11.5 ~\rm{fb}$,
compared with an expected exclusion of $2.6\,{\rm fb}$ (Fig.~3 of Ref.~\cite{ATLAStalk}), showing an excess of $3.6\,\sigma$ significance~\cite{ATLAStalk}. Using Monte Carlo simulation we estimate the acceptance of the fiducial region for scalar production via gluon fusion to be $\sim 60\%$.
In this case, the Gaussian approximation cannot be used to estimate of the signal strength, as is clear also from the very large value of the observed limit, compared to the expected one.
We therefore parameterize the likelihood with a Poissonian function, requiring the correct observed exclusion limit and local significance to be reproduced, resulting in
\begin{equation}
\mu_{13\,{\rm TeV}}^{\rm ATLAS} = \sigma(pp \to S)_{13\,{\rm TeV}} \times \mathcal{B}(S \to \gamma\gamma) = 10^{+4}_{-3}\,{\rm fb}\,.
\end{equation}
The CMS search for a diphoton scalar resonance~\cite{CMS:2014onr} performed during the Run-I phase at proton--proton collision energy of 8 TeV, sets a $95\%$ CL observed upper limit of $\sigma(p p \to S)_{\rm{8\,TeV}} \times \mathcal{B}(S\to\gamma\gamma) < 1.32$~fb, with an expected limit of 0.69~fb (Fig.~10 of Ref.~\cite{CMS:2014onr}), implying that a $\sim 2 \sigma$ excess was observed by CMS already at Run-I.
The analogous ATLAS search~\cite{Aad:2015mna} reported an observed upper limit on the RS graviton production cross section times branching ratio of $< 2.8$~fb at $95\%$ CL, with an expected one of 2.2 fb (Fig.~4 of Ref.~\cite{Aad:2015mna}).
We estimate that the limit improves only by a factor $\sim 1.3$ for the scalar resonance case.
Based on the expected and observed exclusion limits, the diphoton signals from 8~TeV searches in the Gaussian approximation, for a narrow-width scalar resonance, are
\be\begin{split}
\mu^{\rm CMS}_{\rm{8 TeV}} &= \sigma(p p \to S)_{\rm{8\,TeV}} \times \mathcal{B}(S\to\gamma\gamma) 
= (0.63\pm 0.35)~\rm{fb}\,, \\
\mu^{\rm ATLAS}_{\rm{8 TeV}} &= \sigma(p p \to S)_{\rm{8\,TeV}} \times \mathcal{B}(S\to\gamma\gamma)
= (0.46\pm 0.85)~\rm{fb}\,.
	\label{eq:CMSaa8TeVbound}
\end{split}\ee
Before presenting any further discussions, we note that there is a simple (and yet general) way to test the compatibility of 8 and 13 TeV measurements, by fitting to a single parameter:
\begin{equation}
\mu_{\rm{13 TeV}} = R_{pp} ~ \mu_{\rm{8 TeV}}\,,
\end{equation}
where $R_{pp}$ depends on the production mechanism of the scalar and, to a good approximation, is given by the ratio of the parton luminosities of the relevant initial states at the two collision energies. For instance, this ratio is $4.7$, $2.5$ and $2.7$ for $750$~GeV scalar produced via gluon fusion ($g g$), vector fusion ($\sum q q + q \bar q$) and associated production ($\sum q \bar q$), respectively~\cite{partonLumi}.
Among these, the most likely possibility to reconcile the measurements at two energies is if the scalar was produced dominantly via gluon fusion. 

\begin{figure}
  \begin{center}
    \includegraphics[width=0.6\textwidth]{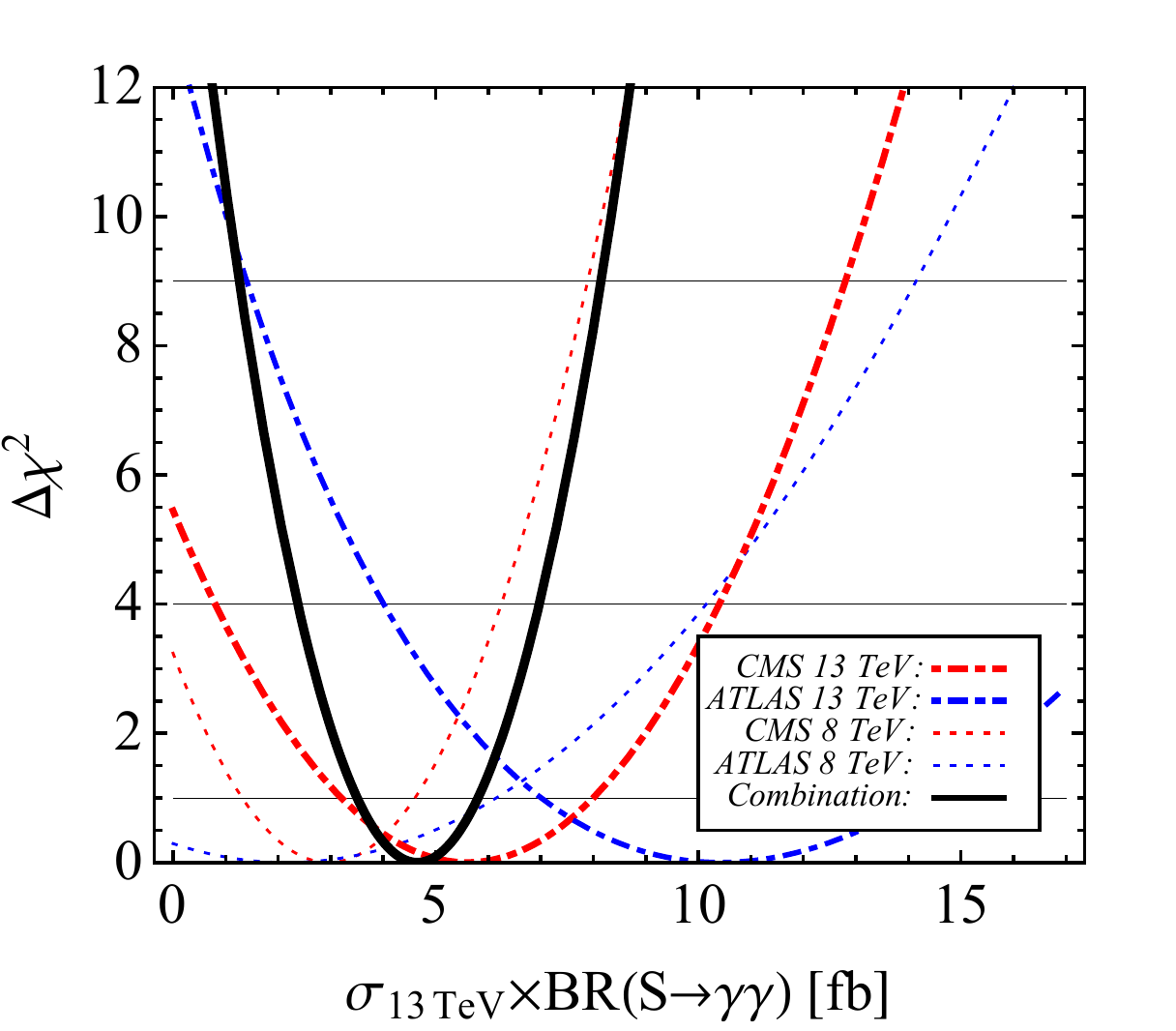}
  \end{center}
\caption{\small\label{fig:combination} Reconstructed likelihoods as a function of $\mu_{\rm{13 TeV}}$ of the diphoton resonance searches at $8$~TeV (dashed) and $13$~TeV (dotted-dashed) by CMS (red) and ATLAS (blue). The combination (solid-black) is obtained when interpreting the resonance as a narrow-width (pseudo)scalar particle produced via gluon fusion.  }
\end{figure}

Finally, when interpreting the resonance as a narrow-width (pseudo-) scalar particle produced via gluon fusion, our combination of 8 and 13 TeV measurements leads to
\begin{equation}
\mu_{\rm{13 TeV}} = (4.6 \pm 1.2)\, \rm{fb}\,,
\label{eq:combination}
\end{equation}
still showing a significant excess over the SM background. Shown in Fig.~\ref{fig:combination} are the individual likelihoods at 8~(13)~TeV in dashed (dotted-dashed) both for CMS (red) and ATLAS (blue), while the combination is shown in solid-black.

Another important experimental input, besides the overall signal yield, is the decay width of the resonance ($\Gamma_S$). The typical diphoton invariant mass resolution of the detector at $750$~GeV is approximately $\sim 10~\GeV$ \cite{Aad:2015mna}. On the one hand, if the width is much smaller than the resolution, as for the SM Higgs, it remains experimentally unmeasurable. On the other hand, for a sizable width a direct measurement from the lineshape is possible. Based on the results presented in \cite{CMStalk,ATLAStalk}, the observed width of the excess is easily compatible with $\Gamma_S \lesssim 40$~GeV. Indeed, the  best-fit in the ATLAS analysis~\cite{ATLAStalk} is obtained for $\Gamma_S/m_S \sim 6\%$ with $3.9~\sigma$ local significance (compared with $3.6~\sigma$ in the narrow-width approximation). It is clear that the present data is not yet conclusive and a precise measurement of the width would require further analysis.
Until then, we study the phenomenological implications for both cases: negligible and sizable width (compared to the resolution).

In the rest of the paper we entertain the possibility that the excess is due to a new scalar (or pseudo-scalar) resonance. In Sec.~\ref{sec:simplemodel} we introduce a simplified effective framework to parametrize its couplings to SM particles, while in Sec.~\ref{sec:constraints} we explore all the relevant experimental constraints from other searches performed at Run-I. In Sec.~\ref{sec:scenarios} we discuss two phenomenologically different scenarios based on the size of the total decay width. Finally, in Sec.~\ref{sec:models} we speculate on possible interpretations of the excess in some concrete new physics models, after which we conclude.

\section{Simplified characterization framework}
\label{sec:simplemodel}

Let us add to the Standard Model (SM) a neutral scalar resonance $S$ with mass $m_S$.
The relevant effective operators for $g g \to S \to \gamma \gamma$ are
\begin{equation}
\cL^{\rm eff} \supset c_{G} \frac{\alpha_s}{12 \pi m_S} S  \, G_{\mu \nu}^a G^{a,\mu\nu} + c_{\gamma \gamma} \frac{\alpha}{4 \pi m_S} S F_{\mu \nu} F^{\mu \nu} ~,
	\label{eq:effLagrMass}
\end{equation}
where $G^{a}_{\mu\nu}$ and $F_{\mu\nu}$ are the $\SU(3)_C$ and $\U(1)_{\rm{QED}}$ field strength tensors, respectively, while $\alpha_s$ and $\alpha$ are the strong and electromagnetic coupling constants. 
Analogously, if the resonance is a pseudo-scalar, the effective Lagrangian describing its interactions with SM particles remains the same as in Eq.~\eqref{eq:effLagrMass} after the substitution
\be
	c_{G,\gamma \gamma} \to \tilde c_{G, \gamma\gamma} \;\;\; \text{and} \;\;\; X_{\mu\nu} X^{\mu\nu} \to \frac{1}{2} \epsilon^{\mu\nu\rho\sigma} X_{\mu\nu} X_{\rho\sigma}~,
	\label{eq:pseudoscalarSub}
\ee
where $(X = G,F)$.

We assume that the dominant production mechanism of the scalar $S$ at hadron collider is gluon fusion ($g g \to S$), induced by $c_{G}$.
As anticipated in the literature, the leading order cross section is expected to receive large higher order QCD correction factor~\cite{Djouadi:2005gi}. We benefit from the precise computations of the SM Higgs boson production at the LHC. In fact, the QCD correction factor in the infinite top mass limit (to match our calculation) is a good approximation of the full top mass dependence even for a relatively heavy Higgs boson~\cite{Djouadi:2005gi}. In the numerical analysis below, we fix $m_S=750$~GeV and estimate the total cross section using the available NNLO QCD  predictions for the SM Higgs~\cite{Djouadi:2005gi,Heinemeyer:2013tqa} and rescaling these for the heavy $m_{\rm top}$ limit (since we assume heavy mediators in the loops generating Eq.~\eqref{eq:effLagrMass}). 
Finally, the production cross sections for a $750$~GeV scalar are
\begin{equation}
\begin{aligned}
	\sigma(p p \to S)_{\rm{8\,TeV}} &= c_{G}^2 \times (12 \pm 1)~\rm{fb}~,\\
	\sigma(p p \to S)_{\rm{13\,TeV}} &= c_{G}^2 \times (55 \pm 6)~\rm{fb}~.
	\label{eq:gg_prod}
\end{aligned}
\end{equation}
For the pseudo-scalar case, we note that the N$^3$LO QCD corrections, computed recently in Ref.~\cite{Ahmed:2015qda} for 13 TeV, provide a K-factor very similar to the one of the scalar \cite{Djouadi:2005gi}, well within the scale uncertainty of the NNLO computation.

The decay width of $S$ into two photons or two gluons following from Eq.~\eqref{eq:effLagrMass} is given by
\ba
	\Gamma(S \to \gamma \gamma) &=&  \frac{m_S}{4 \pi}\left(\frac{\alpha \, c_{\gamma\gamma}}{ 4 \pi} \right)^2 \simeq 2.3 \times 10^{-5}\, c_{\gamma\gamma}^2~\GeV~,
	\label{eq:BrAA} \\
	\Gamma (S \to g g) &=& \frac{2 m_S}{\pi} \left( \frac{\alpha_s c_G}{12 \pi}\right)^2 K_F \simeq 4.1 \times 10^{-3} \,c_G^2~\GeV~,
	\label{eq:BrGG}
\ea
where we include the NLO QCD correction factor $K_F = 1+\frac{67 \alpha_S}{4 \pi}$ from Ref.~\cite{Djouadi:2005gi}, and the running coupling constant at the appropriate scale, $\alpha_s\equiv \alpha_s(m_S)$.
Note that the leading order partial decay widths for the pseudo-scalar case are the same, with the substitution $c_X \to \tilde c_X$.
Since the $c_{\gamma\gamma}$ and $c_{G}$ effective couplings are expected to be generated at loop level ($c_{\gamma\gamma,G} \sim \mathcal{O}(1)$), any tree-level coupling of $S$ to lighter states is typically expected to substantially increase the total decay width to $ \sim \mathcal{O}(\GeV)$, thus strongly reducing the branching fraction in two photons, in which case a larger value of both $c_G$ and $c_{\gamma\gamma}$ is needed to fit the excess. In the Sec.~\ref{sec:largewidth} we quantify this statement.

In order to assess the allowed parameter space for a generic model with a 750 GeV neutral scalar, we parametrize its tree-level couplings to SM particles as follows:
\be
	\cL_{S - \text{SM}} = c_V \frac{S}{m_S} \left( m_Z^2 Z_\mu Z^\mu + 2 m_W^2 W_\mu^+ W^{- \mu} \right)
	+ c_f  \frac{S}{m_S} m_f \bar{f} f + \frac{S}{m_S} \left( c_{h\partial} \partial_\mu h \partial^\mu h - c_{hm} \frac{m_h^2}{2} h^2 \right) ~.
	\label{eq:LagrTree}
\ee
The partial decay widths which follow from these couplings are
\ba
\label{eq:tree-dec}
	\Gamma(S \to ZZ) &=& \frac{c_V^2 m_S}{32 \pi} \sqrt{1 - \beta_Z} \left( 1 - \beta_Z + \frac{3}{4} \beta_Z^2 \right) \simeq 6.8 \, c_V^2 \, \GeV\,, \nonumber \\
	\Gamma(S \to WW) &=& \frac{c_V^2 m_S}{16 \pi} \sqrt{1 - \beta_W} \left( 1 - \beta_W + \frac{3}{4} \beta_W^2 \right) \simeq 13.9 \, c_V^2 \, \GeV \, , \label{eq:treedecays}  \\
	\Gamma(S \to hh) &=& \frac{m_S}{32 \pi} \left(c_{h\partial} (1-\beta_h / 2) + c_{hm} \frac{\beta_h}{4} \right)^2\! \sqrt{1 - \beta_h} 
	\simeq 6.3 \, (c_{h\partial} + 0.029 c_{hm} )^2\, \GeV\,, \nonumber\\
	\Gamma(S \to f \bar{f}) &=& \frac{c_f^2 N_c}{8 \pi m_S} \bar{m}_f^2 (1 - \beta_f)^{3/2} ( 1 + \Delta_{\rm QCD} ) ~ \stackrel{f = t}{\simeq} ~ 3.3 \, c_t^2 \, \GeV \, ,\nonumber 
\ea
where $\beta_X = 4 m_X^2 / m_S^2$. Also, $\bar{m}_f$ is the $\overline{\rm MS}$ mass that should be evaluated at the scale $m_S$ and $\Delta_{\rm QCD}$ are the higher order QCD corrections \cite{Djouadi:2005gi}. In the $WW$ and $ZZ$ decay formulas given in Eq.~\eqref{eq:treedecays} we  neglected the contributions from loop-induced couplings, since these are expected to be subleading whenever the tree-level couplings are present. The tree-level couplings to top and $W$ induce at 1-loop the effective $S\gamma\gamma$ and $Sgg$ interactions. Their contribution to the decay widths and production cross section can be obtained with the substitution
\be\begin{split}
	c_{\gamma\gamma}^2 &\to \left| c_{\gamma\gamma} + (-0.74 + 0.94 i ) c_V + (0.30 - 0.74 i) c_t N_c Q_{t}^2 \right|^2\,,\\
	c_G^2 &\to \left| c_{G} +\frac{3}{4} (0.59 + 1.5 i ) c_t \right|^2\,.
	\label{eq:loopmatch}
\end{split}\ee
Let us analyze the question if the excess can be accommodated by top coupling only, without invoking further contributions to $c_G$ and $c_{\gamma  \gamma}$. Using the expressions derived in this section, we find that the required coupling is $c_t \gtrsim 50$. This, however, would imply an unphysical partial decay width $\Gamma(S\to t \bar t)\sim8~\TeV$. We also note that the situation cannot be ameliorated by considering non-zero values of $c_V$, nor by introducing a large coupling $c_b$ to the bottom quark as well.
We therefore exclude this possibility and conclude that new colored and charged particles inside the loop are necessary in order to explain the excess. As we will see later on, these particles are expected to be light in order to accommodate the excess, potentially within the reach of LHC.

In addition to the decay channels listed in Eq.~\eqref{eq:treedecays}, we also consider a possible invisible decay width $\Gamma_{\rm inv}$. This could be due to decays into dark matter particles, or into particles which escape detection, or for which no experimental bound is present.

\section{Experimental constraints}
\label{sec:constraints}

\begin{table}[t]
\begin{center}
\begin{tabular}{c|c|c} \hline
 Channel & CMS bound [fb] & ATLAS bound [fb] \\
	$\gamma\gamma$ & 1.3~\cite{CMS:2014onr} & 2.2~\cite{Aad:2015mna} \\
	$gg$ & $1.8 \times 10^3$~\cite{CMS:2015neg} & -- \\
	$ZZ$ & 27~\cite{Khachatryan:2015cwa} & 12~\cite{Aad:2015kna} \\
	$Z \gamma$ & -- & 6~\cite{Aad:2014fha} \\
	$WW$ & 220~\cite{Khachatryan:2015cwa} & 38~\cite{Aad:2015agg} \\
	$hh$ & 52~\cite{Khachatryan:2015yea} & 35~\cite{Aad:2015xja} \\
	$t\bar{t}$ & $6\times 10^2$~\cite{CMS:lhr} &  $7\times 10^2$~\cite{Aad:2015fna} \\
	Inv. & -- & $\sim$ $3\times 10^3$~\cite{Aad:2015zva} \\
 \end{tabular}
\caption{\label{tab:constraints} Summary table of LHC Run-I resonance searches showing observed 95\%~CL exclusion limits on $\sigma(g g \to S)\times \mathcal{B}(S\to X X)$ for various decay channels $X X$.}
\end{center}
\end{table}

The framework introduced above can be employed to analyze other potentially relevant experimental constraints in a model-independent way.
In Table~\ref{tab:constraints} we summarize the LHC Run-I limits on $\sigma \times \mathcal{B}$ for a given decay channel for a narrow $750$~GeV neutral scalar (pseudo-scalar) resonance ($\Gamma_S / m_S <$~few~\%). The extended discussion is given below.

The CMS search for a dijet resonance~\cite{CMS:2015neg} at 8 TeV and 18.8~fb$^{-1}$, optimized for the mass window between $500$--$800$~GeV, imposes a $95\%$ CL upper limit on the production of a RS graviton decaying to $g g$:
\be
	\sigma(p p \to X)_{\rm{8\,TeV}} \times \mathcal{B}(X\to g g)\times A < 1.8~\text{pb}~,
	\label{ggbound}
\ee
where $A$ is the acceptance. We conservatively assume $A=1$.

The ATLAS~\cite{Aad:2015kna,Aad:2015agg} and CMS~\cite{Khachatryan:2015cwa} searches for a scalar resonance decaying to $Z Z$ and $W W$ with the full data set, combining all the relevant $Z$ and $W$ decay channels, impose a $95\%$~CL upper limits of
\begin{align}
&\sigma(p p \to S)_{\rm{8\,TeV}} \times \mathcal{B}(S\to Z Z) < 12\,{\rm fb}\,_{\rm(ATLAS)} \,,~ 27\,{\rm fb}\,_{\rm(CMS)}\,,\label{ZZbound}\\
&\sigma(p p \to S)_{\rm{8\,TeV}} \times \mathcal{B}(S\to W W) < 38\,{\rm fb}\,_{\rm(ATLAS)} \,,~ 220\,{\rm fb}\,_{\rm(CMS)}\,.\label{WWbound}
\end{align}
On the other hand, the ATLAS collaboration~\cite{Aad:2014fha} has performed a search for resonance decaying to $\gamma$ and $Z(\to \ell^+ \ell^-) $. From Monte Carlo simulations we estimate that $\sim 70\%$ of all events fall into the fiducial region, and we interpret the search as a $95\%$~CL  upper limit on the inclusive $\sigma \times \mathcal{B}$,
\begin{equation}
\sigma(p p \to S)_{\rm{8\,TeV}} \times \mathcal{B}(S\to Z \gamma) \lesssim 6\,{\rm fb}\,_{\rm(ATLAS)}~.
\end{equation}

The decay into a Higgs boson pair is also subject to constraints. Several searches have been performed during the LHC Run-I by ATLAS \cite{Aad:2015xja} and CMS \cite{Khachatryan:2015yea}; for a resonance mass of 750 GeV, the most stringent constraints come from the $4b$ channel, where each of the 125 GeV Higgs bosons decays into a $b\bar b$ pair. One has
\be
	\sigma(pp\to X)_{8\,{\rm TeV}}\times \mathcal{B}(X\to hh) < 35\,{\rm fb}\,_{\rm (ATLAS)}\,,~ 52\, {\rm fb}\,_{\rm (CMS)}\,.
\ee
This constrains the $\mathcal{B}(S\to hh)$ at a level similar to the branching ratio into vector bosons, i.e. a few $10^{-2}$ for a production cross cross sectionsection of the order of a few picobarn.

Decays into quarks are experimentally less constrained. In particular, ATLAS \cite{Aad:2015fna} and CMS \cite{CMS:lhr} searches for resonance decaying into a pair of top quarks result in the following constraints:
\be
	\sigma(pp\to X)_{8\,{\rm TeV}}\times \mathcal{B}(X\to t\bar t) < 0.7\,{\rm pb}\,_{\rm (ATLAS)}\,,~0.6\,{\rm pb}\,_{\rm (CMS)}\,.
\ee
Other fermionic channels, such as $b\bar b$, $\tau^+\tau^-$, or light quarks and leptons, are less relevant if one assumes Yukawa-like couplings to the new resonance as in Eq.~\eqref{eq:LagrTree}.

An interesting possibility is that $S$ predominantly decays to invisible particles that might constitute (part of) the observed dark matter in the universe. This scenario, on the other hand, might lead to sizable mono-jet  (missing energy plus jet) signatures at the LHC. The present ATLAS search~\cite{Aad:2015zva} sets a bound on $\sigma(pp \to H) \times \mathcal{B}(H \to \text{inv})$ for a heavy Higgs-like particle, but only up to a mass of 300 GeV. We perform a Monte Carlo simulation of the signal for $750$~GeV applying the same cuts as in the analysis. We find that the acceptance times efficiency  improves by a factor of $\sim 3$ with respect to $300$~GeV case. We use this to estimate the bound on a 750 GeV scalar
\be
	\sigma(pp\to S)_{8\,{\rm TeV}}\times \mathcal{B}(S\to \text{inv}) \lesssim 3~\rm{pb}.
\ee

\begin{figure}
  \begin{center}
    \includegraphics[width=0.5\textwidth]{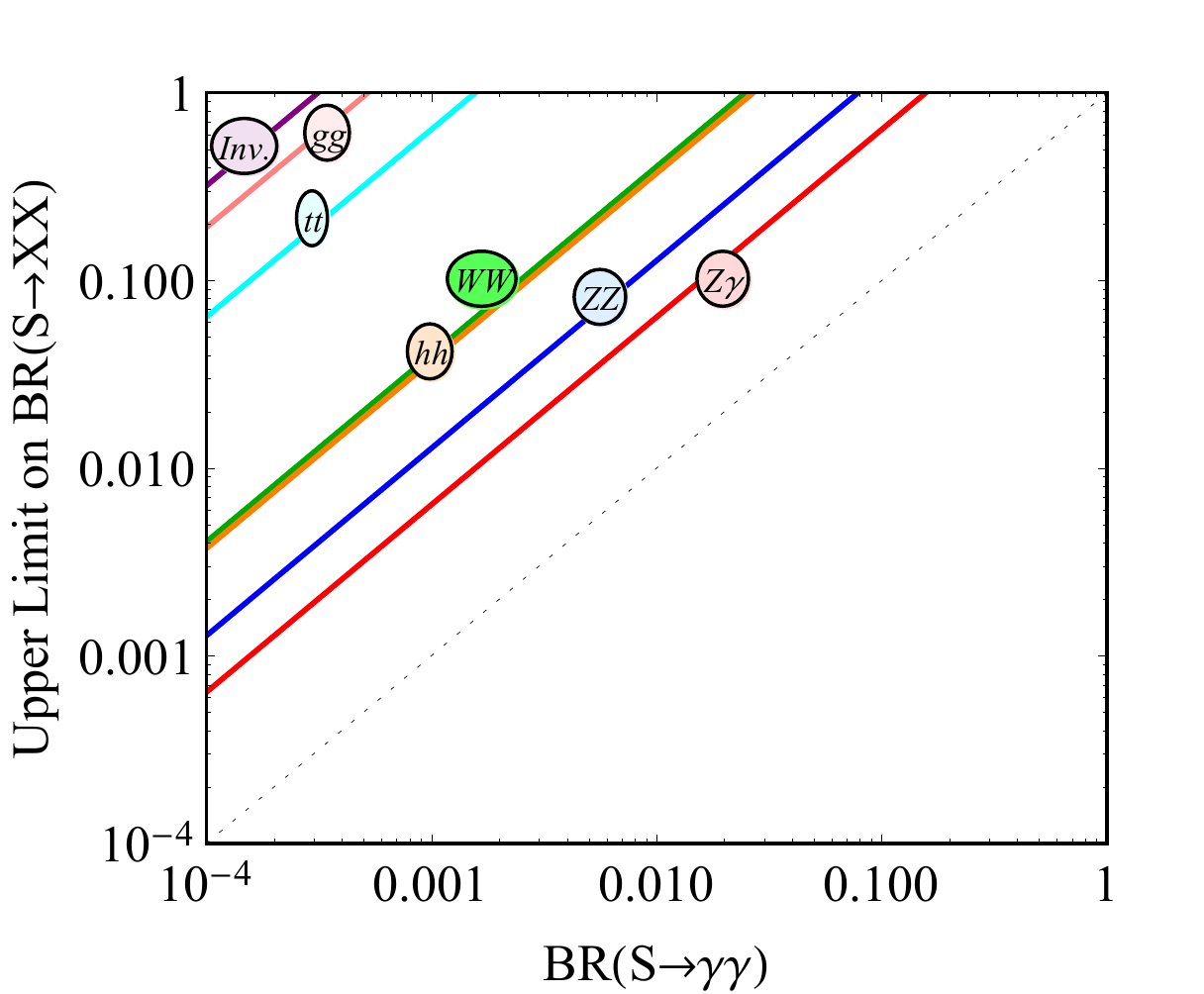}
  \end{center}
\caption{\small\label{fig:br-limits} Upper limits at $95\%$ CL on $\mathcal{B}(S\to X X)$, where $X X=Z\gamma$, $Z Z$, $W W$, $h h$, $t \bar{t}$, $g g$ and invisible, based on LHC Run-I resonance searches, and assuming diphoton signal from Eq.~\eqref{eq:combination}. }
\end{figure}

Negative results from $8$~TeV resonance searches, together with the positive signal in $\gamma \gamma$ channel from Eq.~\eqref{eq:combination}, imply model-independent upper limits on $\mathcal{B}(S\to X X)/\mathcal{B}(S\to \gamma \gamma)$, for a given channel $X X$. These are illustrated in Fig.~\ref{fig:br-limits} for $X X=Z\gamma$, $Z Z$, $W W$, $h h$, $t \bar{t}$, $g g$ and invisible.

\section{Phenomenological scenarios}
\label{sec:scenarios}

With this experimental results in mind, we face two different classes of scenarios, depending on whether or not the resonance has extra tree-level decay channels, that would lead to a sizable decay width. Since this is an open issue in the present analyses~\cite{CMStalk,ATLAStalk}, we study both cases separately.
\begin{itemize}

	\item {\bf Only loop-induced decays:} The dominant decay channels of $S$ are into the SM gauge bosons ($g g$, $\gamma \gamma$, $Z \gamma$, $Z Z$,pseudo-scalar and $W^+ W^-$) through loops of heavy states of the new sector, described effectively by
\begin{equation}
\cL \supset  c_G \frac{\alpha_s}{12 \pi m_S} S  \, G_{\mu \nu}^a G^{a,\mu\nu} +\frac{\alpha}{4 \pi m_S} S ~(c_W W_{\mu\nu}^i W^{i,\mu\nu} + c_B B_{\mu \nu} B^{\mu\nu} ) ~,
	\label{eq:effLagrGauge}
\end{equation}
in $\SU(2)_L\times \U(1)_Y$ invariant way. In terms of the mass eigenstates, the relevant part of the above Lagrangian reads
\begin{equation}
\cL \supset \frac{\alpha}{2 \pi m_S} S ~ ( \frac{c_{\gamma \gamma}}{2} F_{\mu \nu} F^{\mu \nu}+c_{z \gamma} Z_{\mu \nu} F^{\mu \nu} +\frac{c_{z z}}{2} Z_{\mu \nu} Z^{\mu \nu} + c_{w w} W_{\mu\nu}^+ W^{-\mu\nu} ) ~,
	\label{eq:effLagrMass2}
\end{equation}
where
\begin{equation}
\begin{aligned}
& c_{w w} = c_W~, &
 c_{\gamma \gamma} = c_B \cos^2 \theta_W +c_W\sin^2 \theta_W~,\\
& c_{z z} = c_B \sin^2 \theta_W +c_W\cos^2 \theta_W~, \qquad &
 c_{z \gamma} = (c_W-c_B) \sin \theta_W \cos \theta_W~.
 \label{eq:transl_mass_gauge}
\end{aligned}
\end{equation}
Analogous operators can be written in the pseudo-scalar case using the substitutions in Eq.~\eqref{eq:pseudoscalarSub},
\begin{equation}
\cL^{\rm pseudo} \supset  \tilde c_G \frac{\alpha_s}{12 \pi m_S} S  \, G_{\mu \nu}^a \tilde G^{a,\mu\nu} +\frac{\alpha}{4 \pi m_S} S \left(\tilde c_W W_{\mu\nu}^i \tilde W^{i,\mu\nu} + \tilde c_B B_{\mu \nu} \tilde B^{\mu\nu} \right) ~.
	\label{eq:effLagrGaugePseudo}
\end{equation}
In this scenario, the relevant phenomenology depends entirely on $c_G$, $c_W$ and $c_B$ (or $\tilde c_G$, $\tilde c_W$ and $\tilde c_B$).

	\item {\bf Sizable extra decay channels:} $S$ is allowed to couple at tree level to other lighter particles (either SM or BSM ones) and, therefore, $\Gamma_S$ is expected to be dominated by extra channels, rendering it largely independent on $c_{G}$ and $c_{\gamma\gamma}$. In this case, the phenomenological parameters relevant to the observed excess are $c_{G}$, $c_{\gamma\gamma}$, and $\Gamma_S$.
\end{itemize}

\subsection{Phenomenology of the ``loop-only'' scenario}
\label{sec:loop}

The loop-induced decay channels generated by the effective couplings in Eq.~\eqref{eq:effLagrGauge} are dominant only if any tree-level couplings to lighter states are strongly suppressed.
Starting with Eq.~\eqref{eq:effLagrGauge}, we compute the partial decay widths for $S\to V V$, where $VV=\gamma\gamma$, $Z Z$, $W W$, $Z \gamma$ and $g g$, as a function of $c_G$, $c_B$, and  $c_W$.
The total width is then simply given by the sum of the partial widths,
\be
	\Gamma_S = \Gamma_S^{loop}(c_G, c_W, c_B)=\sum_{VV} \Gamma_{VV}
	\simeq (4.1 c_G^2 + 0.064 c_W^2 + 0.022 c_B^2 
	) \times 10^{-3}~\text{GeV}~,
\ee
for $m_S = 750$ GeV. The branching ratio into two photons, $\mathcal{B}(S \to \gamma\gamma)$, is also a function of the same three coefficients only.

\begin{figure}
  \begin{center}
    \includegraphics[width=0.44\textwidth]{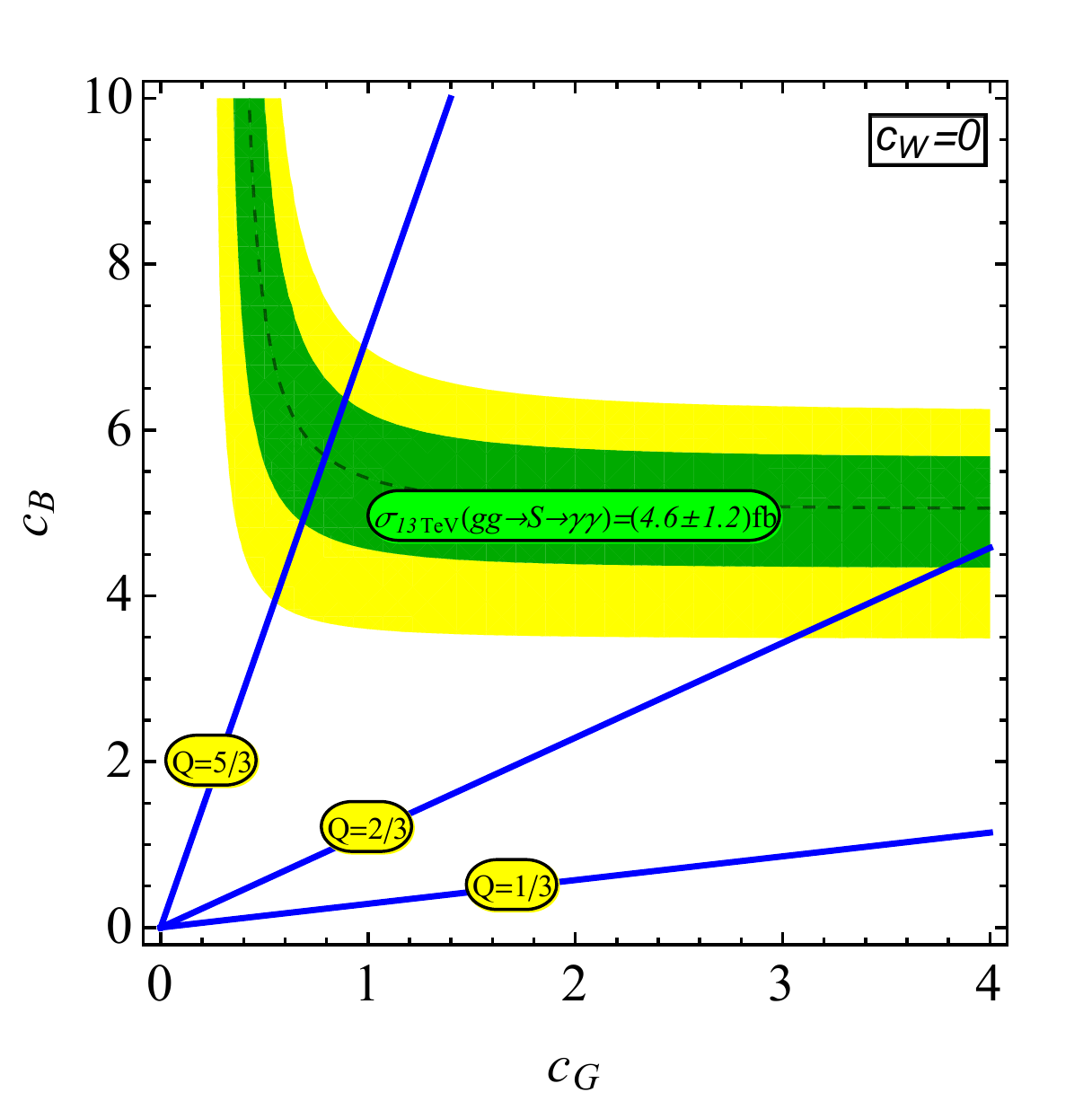}
     \includegraphics[width=0.44\textwidth]{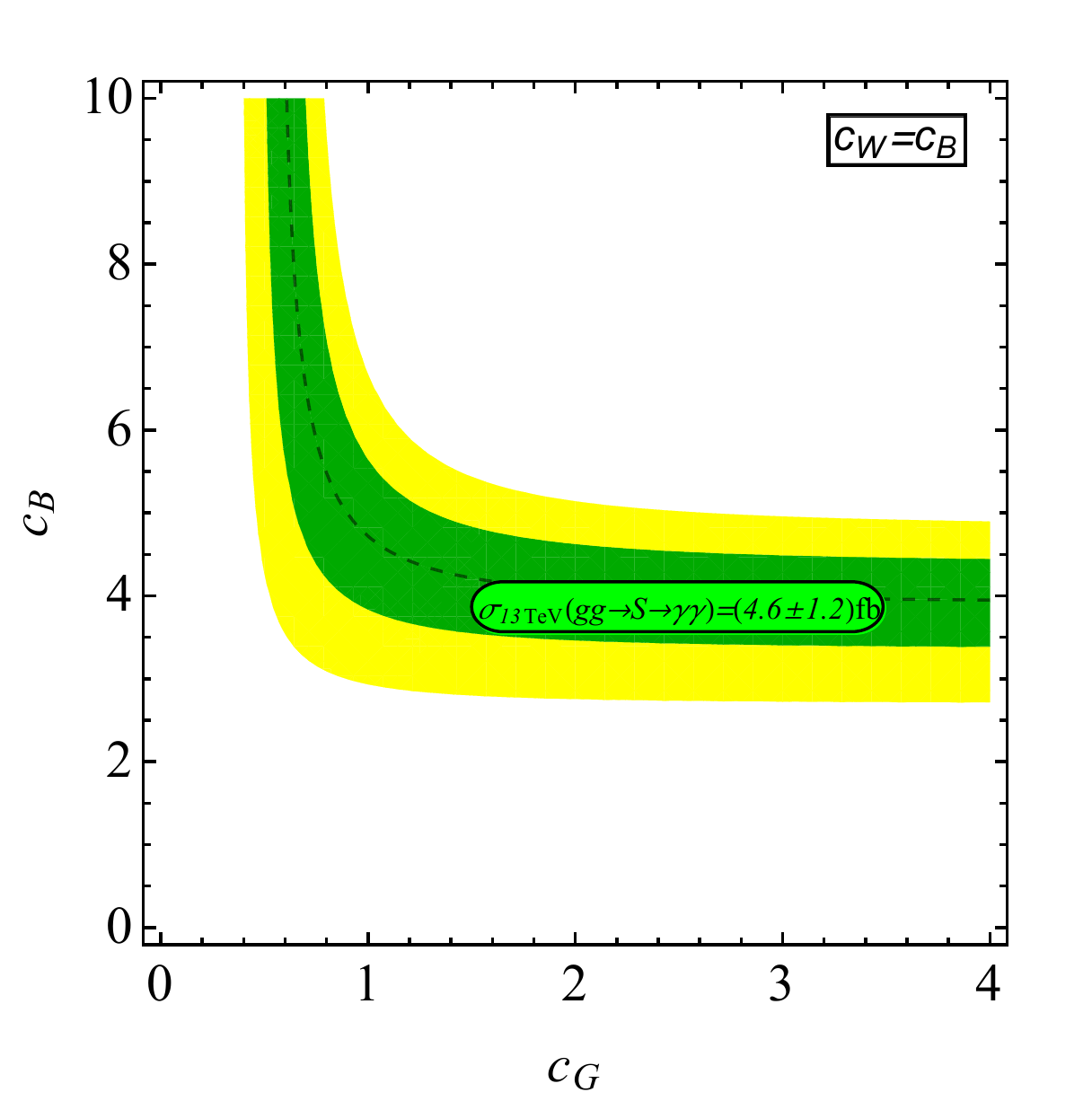}
  \end{center}
\caption{\small\label{fig:excess} Preferred region at $68\%$ (green) and $95\%$ (yellow) CL by the diphoton excess from Eq.~\eqref{eq:combination} in the $(c_G, c_B)$ plane while setting $c_W=0$ (upper plot) and  $c_W=c_B$ (lower plot)  for the ``loop-only'' scenario. Shown in blue are the predictions from vector-like quark model. See text for details.}
\end{figure}

In Fig.~\ref{fig:excess}, we show $68$ and $95\%$ CL preferred regions from the combined diphoton signal in Eq.~\eqref{eq:combination}, translated to the $(c_G, c_B)$ plane, while assuming $c_W=0$ (upper plot) and $c_W=c_B$ (lower plot). Interestingly, the excess can easily be accommodated for typical values of effective couplings generated at one loop.
The other constraints discussed in Sec.~\ref{sec:constraints} are instead not relevant in this scenario for the values of $c_G$, $c_B$, and $c_W$ necessary to fit the excess in Fig.~\ref{fig:excess}. In fact, the dijet limit from Eq.~\eqref{ggbound} sets a loose bound on the effective $S g g$ coupling, $|c_G| \lesssim 12$. On the other hand, this scenario also predicts a correlated signal in $Z Z$, $Z \gamma$, and $W^+ W^-$ channels:
\begin{equation}
\begin{aligned}
\frac{\sigma(p p \to S\to Z \gamma)}{\sigma(p p \to S\to \gamma \gamma)} &= \frac{2 (1- R_{WB})^2 \tan^2 \theta_W }{(1+R_{WB} \tan^2 \theta_W )^2}~, \\
\frac{\sigma(p p \to S\to Z Z)}{\sigma(p p \to S\to \gamma \gamma)} &= \frac{(\tan^2 \theta_W+R_{WB})^2}{(1+R_{WB} \tan^2 \theta_W )^2}~,\\
\frac{\sigma(p p \to S\to W^+ W^-)}{\sigma(p p \to S\to \gamma \gamma)} &=\frac{2 R_{WB}^2}{(\cos^2 \theta_W +R_{WB} \sin^2 \theta_W )^2}~,
\end{aligned}
\end{equation}
where $R_{WB} = c_W/c_B$. The best present limit is due to the $Z \gamma$ channel (see Fig.~\ref{fig:br-limits}) with a lower limit $R_{WB} \gtrsim -1.7$ and a loose upper limit. The correlation induced via the single parameter $R_{WB}$ is a striking prediction for future searches at LHC Run-II.
In Sec.~\ref{sec:models} we discuss the interpretation of the excess in some concrete models which fall into this class.

\subsection{Phenomenology of the ``extra-width'' scenario}\label{wide}
\label{sec:largewidth}

In this section we entertain the possibility that the decay width of the new scalar is dominated by other decay channels than the loop generated ones described previously, in which case the total decay width $\Gamma_S$ becomes an independent free parameter.
This implies that the branching fraction in two photons, $\mathcal{B}(S \to \gamma\gamma)$, is a function of $c_{\gamma\gamma}$ and $\Gamma_S$ only.

Within this class of scenarios, two possibilities could be realized:
\begin{itemize}
\item either the width is bigger than the experimental resolution ($\Gamma_S \gtrsim 10~\GeV$) and could then be measured directly,
\item or it is still smaller than the resolution, $ \Gamma_S \lesssim 10~\GeV$, but larger than the ``loop-only'' contribution, $\Gamma_S^{loop} < \Gamma_S$.
\end{itemize}

\begin{figure}
  \begin{center}
    \includegraphics[width=0.44\textwidth]{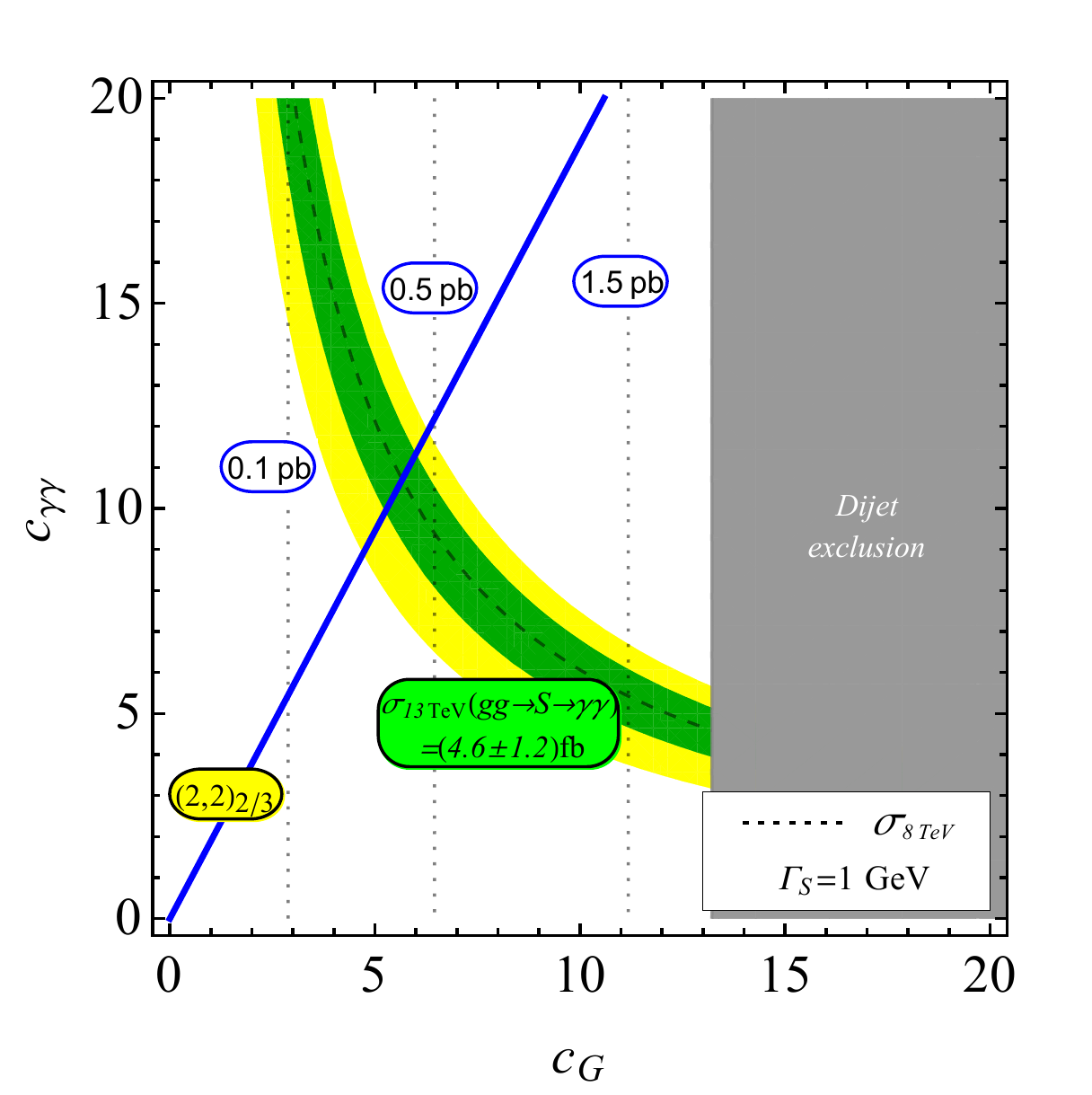}
     \includegraphics[width=0.44\textwidth]{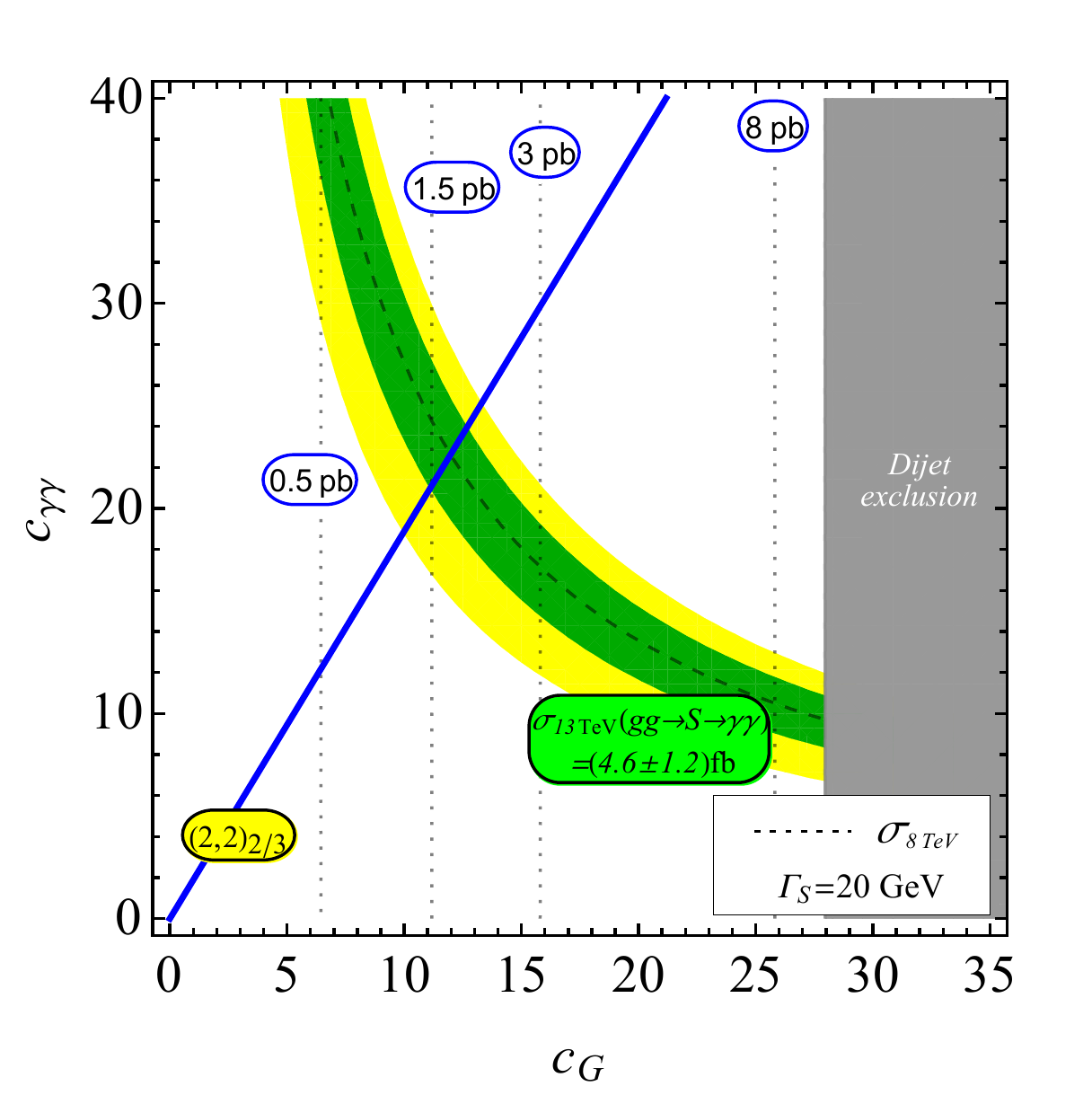}
  \end{center}
\caption{\small\label{fig:largewidth} Preferred region at $68\%$ (green) and $95\%$ (yellow) CL by the diphoton excess from Eq.~\eqref{eq:combination} and constraints from dijet production in Eq.~\eqref{ggbound} in $(c_G, c_{\gamma \gamma})$ plane for $\Gamma_S =1$~GeV (upper plot) and $\Gamma_S =20$~GeV (lower plot). Shown in black-dotted is the gluon fusion production cross section at 8~TeV. The blue line is the prediction for a composite vector-like fermion bidoublet  $({\bf 2},{\bf 2})_{2/3}$. See text for details.}
\end{figure}

As a benchmark point, we assume the total decay width to be either $\Gamma_S =20$~GeV or $\Gamma_S =1$~GeV, and repeat the survey of the relevant phenomenological constraints. 
In this case, the diphoton signal strength at 13 TeV is
\be
	\mu_{\rm{13\,TeV}} = \sigma(pp \to S) \times \mathcal{B}(S \to \gamma\gamma)_{13\,{\rm TeV}}
	\simeq 6.3 \times 10^{-5} \left( \frac{20~\GeV}{\Gamma_S} \right) c_G^2 c_{\gamma\gamma}^2~ \text{fb}~,
\ee
where one can notice the simple scaling with $\Gamma_S$.  We show in Fig.~\ref{fig:largewidth} in green (yellow) the 68~\%~(95~\%)~CL region preferred by the combined fit to diphoton resonance searches from Eq.~\eqref{eq:combination}. The main difference with respect to the previous case, is that the $c_{G,\gamma\gamma}$ couplings have to be large in order to fit the diphoton excess. This should also be confronted with the dijet bound from Eq.~\eqref{ggbound},
\be
\sigma(p p \to S)_{\rm{8\,TeV}} \times \mathcal{B}(S \to g g) \simeq
 2.4 \times 10^{-3} \left( \frac{20~\GeV}{\Gamma_S} \right) c_{G}^4 ~\text{fb} \lesssim 1.8 \times 10^3~\text{fb}~,
\ee
which corresponds to a constraint $|c_G| \lesssim 13$ if $\Gamma_S = 1$ GeV as shown in Fig.~\ref{fig:largewidth} (top) and $|c_G| \lesssim 28$ if $\Gamma_S = 20$ GeV, as shown in Fig.~\ref{fig:largewidth} (bottom).

According to Eq.~\eqref{eq:gg_prod}, the large values of $c_G$ enhance the production cross section, therefore it is important and non-trivial to point out to which final states $S$ is allowed to decay and compare with the present experimental constraints discussed in Sec.~\ref{sec:constraints}. 
The total production cross section $\sigma(p p \to S)$ isolines at $8$~TeV are shown with black-dashed vertical lines in Fig.~\ref{fig:largewidth}.
The 95\% CL constraints on tree-level decay channels from 8 TeV resonance searches are
\be
\begin{split}
	\sigma(p p \to S)_{\rm{8\,TeV}} \times \mathcal{B}(S\to Z Z) & \simeq 4.0 \left( \frac{20~\GeV}{\Gamma_S} \right) c_G^2 c_{V}^2 \text{fb} < 12 \rm{fb}, \\
	\sigma(p p \to S)_{\rm{8\,TeV}} \times \mathcal{B}(S\to W W) & \simeq 8.2 \left( \frac{20~\GeV}{\Gamma_S} \right) c_G^2 c_{V}^2 \text{fb} < 38 \rm{fb}, \\
	\sigma(p p \to S)_{\rm{8\,TeV}} \times \mathcal{B}(S\to h h ) & \simeq 3.7 \left( \frac{20~\GeV}{\Gamma_S} \right) c_G^2  (c_{h\partial} + 0.029 c_{hm} )^2 \text{fb} < 35 \rm{fb}, \\
	\sigma(p p \to S)_{\rm{8\,TeV}} \times \mathcal{B}(S\to t \bar t) & \simeq 2.0 \left( \frac{20~\GeV}{\Gamma_S} \right) c_G^2 c_t^2 \text{fb} < 0.6\rm{pb}.
\end{split}
\ee
Comparing these bounds (Table~\ref{tab:constraints}) with the 8~TeV cross section isolines from Fig.~\ref{fig:largewidth}, one can deduce the size of allowed extra decays. A general conclusion is that a large width of $\Gamma_S \sim 20~\GeV$ can be accounted only by decays in $t \bar t$ or invisible, unless extremely large values of $c_{\gamma\gamma}$ are invoked.

\section{Explicit models and insights}
\label{sec:models}

In this section we provide interpretations of the generic scenarios described above, in the context of some more concrete models.

\subsection{Scalar singlet with loop decays only}
\label{sec:narrowscalar}

In this class of models the scalar resonance is a ${\rm SU(2)_L}\times{\rm U(1)_Y}$ singlet, which decays predominantly in SM gauge bosons via the couplings in Eq.~\eqref{eq:effLagrGauge}.
This feature can be naturally achieved by assuming that $S$ is the lightest state of a new heavy sector, coupled to the SM only via the SM gauge interactions, and that it does not participate at tree level to the electroweak symmetry breaking, i.e. there is no mixing with the SM Higgs boson.
In this case the operators in Eq.~\eqref{eq:effLagrGauge} are generated via loops of (colored and charged) heavy states of the new sector.

Given that the main phenomenological constraints concern only the $gg \to S \to \gamma\gamma$ process, the relevant properties of the new particles inside the loop are the spin and the quantum numbers under $\SU(3)_C$ and $\U(1)_{\rm{QED}}$.
As a simple benchmark scenario, we assume $S$ is coupled to a set of vector-like heavy fermions $\Psi_i$, triplets (or singlets) of $\text{SU}(3)_C$ with electric charge $Q_i$ and mass $M_i$, via marginal (Yukawa type) operators, $\cL_{Yuk} \supset - g_i^* S \bar{\Psi}_i \Psi_i$. At one loop they generate the necessary couplings of $S$ with gluon and photon pairs. In the limit of heavy fermions ($\tau_{\Psi_i} = m_S^2 / 4 M^2_i \ll 1$), we can match the $c_G$, $c_{\gamma\gamma}$ and $c_{Z \gamma}$  coefficients to the model parameters,
\be
\begin{aligned}
	c_G &= \sum_{i\in \text{triplets}} g^*_i \frac{m_S}{M_i}~,\\
	c_{\gamma\gamma} &= \frac{2}{3} \sum_i  g^*_i \frac{m_S}{M_i}~N^c_i Q_i^2~,\\
	c_{Z \gamma} &= \frac{2}{3} \sum_i  g^*_i \frac{m_S}{M_i}~ N^c_i Q_i \frac{T_i^3-Q_i \sin^2 \theta_W}{\cos \theta_W \sin{\theta_W}} ~,
\end{aligned}
\label{eq:matching}
\ee
where $N_i^c = 3~(1)$ for the color triplets (singlets) and $T_i^3$ is the $\SU(2)_L$ isospin quantum number. For completeness, $c_{Z \gamma}$ parameterizes the effective scalar coupling to $Z \gamma$.
Using the relations in Eq.~\eqref{eq:transl_mass_gauge}, the matching in Eq.~\eqref{eq:matching} can easily be translated to $c_B$ and $c_W$. 

Based on the analysis in Sec.~\ref{sec:loop}, the ``loop-only'' scenario points to relatively small effective couplings $c_G$, $c_B$ and $c_W$, which can in principle be due to a single (or few) particle(s) in the loop.
For example, consider a single vector-like quark representation which is a singlet under $\SU(2)_L$ and has electric charge $Q_f = Y_f$:
\begin{equation}
\textrm{Vector-like quark }({\bf 3},1,Q_f) :
~~ c_G =  g^* \frac{m_S}{M_f},\quad c_B= \frac{2 Q_f^2}{\cos^2 \theta_W} c_G, \quad c_W=0~.
\end{equation}
In Fig.~\ref{fig:excess} (top), we show in solid-blue the predictions for electric charges $5/3$, $2/3$ and $1/3$. The first two, in particular, can nicely explain the observed excess for reasonable values of $g^* \sim \mathcal{O}(1)$ and $M_f \sim \mathcal{O}$(TeV). If this is correct, one can expect vector-like quark signatures to show up at the LHC.

\subsection{Singlet mixed with the Higgs}
The previous scenario can be generalized allowing also for other decay channels of the singlet into SM particles, which can arise at a renormalizable level through mixing with the Higgs. Indeed, the cubic term $S|H|^2$ will in general be present in the scalar potential, if no particular assumption is made in order to suppress it, and it will give rise to a mass mixing between the two CP even states after electroweak symmetry breaking. A situation of this kind can be found in the context of many most natural extensions of the SM.

The effective Lagrangian for the singlet is then given by Eq.~(\ref{eq:LagrTree}), with
\begin{equation}
c_V = c_f = \frac{m_S}{v}\sin\theta, \qquad\qquad c_{h\partial} = 0,
\end{equation}
where $\theta$ is the singlet--Higgs mixing angle.
For sufficiently high masses, the main decay widths of $S$ are into $W$, $Z$, and Higgs bosons, in an approximate ratio
\begin{equation}\label{decay_singlet}
\Gamma(S\to WW) = 2\Gamma(S\to ZZ) \simeq 2\Gamma(S\to hh)\propto \sin^2\theta,
\end{equation}
dictated by the equivalence theorem. The exact value of $\Gamma(S\to hh)$, which is determined by $c_{hm}$, depends on the details of the scalar potential; see e.g.~\cite{Buttazzo:2015bka}.

If no other relevant decay modes are present, the branching ratios into $W^+W^-$ and $ZZ$ are close to 0.5 and 0.25, respectively.
The total width in this case is $\Gamma_S = \sin^2\theta\, \Gamma_{\rm SM}$, where $\Gamma_{\rm SM} \simeq 250$ GeV is the width of a SM-like Higgs of 750 GeV. One can then see from Fig.~\ref{fig:br-limits} that, due to the bound on $S\to ZZ$, a branching ratio into $\gamma\gamma$ of at least about 2\% is needed in order to reproduce the observed signal if $\mathcal{B}(S\to Z Z) \sim 25\%$. 
Independently of the total decay width, this can be recast as a bound on the mixing angle $\sin \theta \lesssim 2 \times 10^{-3} c_{\gamma \gamma}$.
The strongest experimental constraint on the mixing angle from the LHC Run-I Higgs coupling analysis, $\sin^2\theta < 0.2$~\cite{ATLAS_CMS}, is several orders of magnitude weaker for physically motivated values of $c_{\gamma\gamma}$.

\subsection{Pseudo-scalar singlet}
\label{sec:pseudoscalar}

Similar couplings to photons and gluons can be generated also for a pseudo-scalar singlet $S$, with a coupling to the heavy fermions of the type $\cL_{Yuk} \supset - \tilde{g}_i^* S \bar{\Psi}_i i \gamma_5 \Psi_i$.
In this case, the matching to $\tilde{c}_G$ and $\tilde c_{\gamma\gamma}$ is similar to Eq.~\eqref{eq:matching} with $g^*_i \to 3/2 \tilde{g}_i^*$.
A nice feature of the pseudo-scalar scenario is that the CP symmetry automatically forbids a mixing of this particle with the SM Higgs, thus also forbidding tree-level decays to SM gauge bosons.
Axion-like particles fall in this class of models, an early analysis of collider bounds in this context can be found e.g. in Ref.~\cite{Jaeckel:2012yz}.

In addition to this, if the (pseudo-scalar) singlet is one of the pseudo-Nambu-Goldstone bosons (pNGB) of a non-minimal composite Higgs scenario in a symmetry-breaking pattern $G/H$, one expects that a Wess--Zumino--Witten term in the low-energy effective theory is generated; see e.g. Refs.~\cite{Kilian:2004pp,Gripaios:2009pe}. This effectively provides couplings of the pseudo-scalar singlet to SM gauge bosons exactly as in Eq.~\eqref{eq:effLagrGaugePseudo}, with the matching 
\be
	\tilde c_B = \frac{n_B}{\cos^2 \theta_W} \frac{m_S}{f}~,\,
	\tilde c_W = \frac{n_W}{\sin^2 \theta_W} \frac{m_S}{f}~,\,
	\tilde c_G = 3 n_G \frac{m_S}{f}~,
\ee
where $f$ is the scale of the spontaneous symmetry breaking in the strong sector and $n_{G,W,B}$ are $O(1)$ coefficients which depend on the symmetries and fermion content of the underlying UV theory. Using Eq.~\eqref{eq:transl_mass_gauge} for the pseudo-scalar one obtains $\tilde c_{\gamma\gamma} = (n_B + n_W) m_S / f$.
This contribution, together with those from loops of heavy fermions discussed above, could easily match the observed excess for $O(1)$ values of the $n_{B,W,G}$ parameters (see Fig.~\ref{fig:excess}).
In this context, measuring the $n_{B,W,G}$ parameters could offer an insight into the UV structure of the strong sector \cite{Gripaios:2009pe}, in the same way as measuring the $\pi^0 \to \gamma\gamma$ decay width offered insights on the structure of QCD.

A simple model which provides a singlet pNGB, as well as a solution to the electroweak naturalness problem, can be found in the context of non-minimal composite Higgs models, for example those based on the spontaneous symmetry-breaking pattern $\SO(6)/\SO(5)$ \cite{Gripaios:2009pe,Frigerio:2012uc,Redi:2012ha,Marzocca:2014msa}. In particular, it has been shown  \cite{Redi:2012ha,Marzocca:2014msa} that a $\sim 750~\GeV$ singlet pNGB can be accommodated in such models.
Even though in this case the UV anomaly is such that $n_G = 0$ and $n_B = - n_W$, \footnote{We thank Michele Frigerio for pointing out a sign error in Ref.~\cite{Gripaios:2009pe}, which propagated to the first version of this paper.} which gives vanishing $\tilde c_{\gamma\gamma}$ and $\tilde c_G$, the necessary non-zero contributions to these coefficients in order to explain the excess can be obtained from loops of SM fermions or vector-like top partners, generically predicted in these setups.

\subsection{Composite resonance}
\label{sec:comp}

The ``extra-width'' scenario points to large effective couplings which, based on perturbativity arguments, require several fermion representations to coherently contribute to $g g \to S$.
Interestingly enough, we find this to be plausible in the context of composite Higgs models if $S$ is a composite scalar singlet resonance of the strong sector. In this example we focus on the minimal composite Higgs model $\SO(5)/\SO(4)$ \cite{Agashe:2004rs}, where such a resonance could play a role in the unitarization of $WW$ scattering at high energy \cite{Contino:2011np}. Its mass is naively expected to be near the strong coupling scale $\Lambda\sim\text{(few) TeV}$, unless it is protected by a symmetry. Nevertheless, the resonance could still be light due to some accidental cancellations or peculiarities of non-perturbative dynamics. For example, in QCD the $\sigma$ meson (or $f_0(500)$) is much lighter than the typical mass scale of the other resonances, even though it is not a pNGB like the pions.

In order to generate sizable values of $c_G$ consider, for example, a color triplet vector-like fermion resonance that transforms as a bidoublet $({\bf 2},{\bf 2})_{2/3}$ under the $\SU(2)_L \times \SU(2)_R \times \U(1)_X$ global symmetry of the strong sector. The four mass eigenstates have electric charges ($Q=T^3_L+T^3_R+X$): $5/3$, $2/3$, $2/3$ and $-1/3$. The composite scalar $S$ couples to the bi-doublets via strong sector coupling $g^*$ as described in Sec.~\ref{sec:narrowscalar}. Using Eq.~\eqref{eq:matching}, we find
\begin{equation}
({\bf 2},{\bf 2})_{2/3} : \quad
c_G = 4 g^* \frac{m_S}{M_f}, \quad  c_{\gamma\gamma} = \frac{17}{9} c_G~, 
\quad c_{Z\gamma} = \left( \frac{1}{2} \cot \theta_W - \frac{25}{18} \tan \theta_W \right) c_G~.
\end{equation}
In Fig.~\ref{fig:largewidth}, we show in solid-blue the correlation in the $(c_G,c_{\gamma\gamma})$ plane. The excess can easily be accounted for reasonable values of the parameters, e.g. for $\Gamma_S = 20~\GeV$, $M_f \sim 1~\TeV$, and $g^* \sim 3$.

Such a composite singlet resonance is expected to couple strongly with the pNGBs of the model, in particular, the Higgs and the longitudinal polarizations of the $W$ and $Z$ bosons. In terms of the four pNGBs $\pi^a$, the coupling reads $\cL_{\rm S,CH} \supset a_S  (\partial_\mu \pi^a)^2 S / f$. In terms of the physical states this interaction corresponds to \cite{Contino:2011np}
\be
	\cL_{\rm S,CH} \supset \frac{2 a_S S}{f} \left( \frac{1}{2} (\partial_\mu h)^2 + \left( \frac{m_Z^2}{2} Z_\mu Z^\mu + m_W^2 W_\mu^+ W^{- \mu} \right) (1 + 2 a_h h + \ldots)  \right)~,
\ee
where $f$ is the scale of the spontaneous breaking of the global symmetry in the composite sector, $f \sim 1 ~\TeV$, and $a_S$ is an $O(1)$ parameter. In particular, notice that the derivative coupling with the Higgs is given as $c_{h \partial} = c_V = a_S \frac{m_S}{f}$. With this matching, and using Eq.~\eqref{eq:tree-dec}, the decay width into the pNGBs ($h$, $W$, and  $Z$) is
\be
	\Gamma(S \to {\rm pNGB}) \simeq 27 \, a_S^2 \frac{m_S^2}{f^2} ~\GeV~.
\ee
The strongest constraint from LHC Run-I resonance searches comes from the $ZZ$ decay channel, Eq.~\eqref{ZZbound}. It imposes a bound of
\be
	\mathcal{B}(S \to {\rm pNGB}) \equiv \Gamma(S \to {\rm pNGB}) / \Gamma_S \lesssim \frac{4.0}{c_G^2} ~.
\ee
For $c_G = 5 \, (10)$, as suggested in Fig.~\ref{fig:largewidth} top (bottom) for the bidoublet representation, this corresponds to $\mathcal{B}(S \to {\rm pNGB}) \lesssim 16 \,(4) \%$. This implies that, for reasonable values of the parameters (i.e. assuming no extremely large contributions to $c_{\gamma\gamma}$), a sizable total width can only be obtained via decays
to other channels, such as $t \bar t$ or invisible. Therefore, one should expect to observe the signal in these channels soon if this scenario is indeed realized in Nature. In other words, the upper limit on $\Gamma(S \to ZZ) / \Gamma(S \to \gamma\gamma)$ from Fig.~\ref{fig:br-limits}, corresponds to $a_S \frac{m_S}{f} < 0.6 \times 10^{-2} c_{\gamma\gamma}$. The fact that the coupling to pNGBs has to be suppressed with respect to the naive expectation $a_S \sim 1$, puts in some tension this scenario as a natural interpretation of the excess.

\subsection{A second doublet: the MSSM and beyond}
\label{sec:mssm}

Extra Higgs bosons below the TeV scale are naturally predicted in supersymmetric models. A prime example is the second doublet of the Minimal Supersymmetric SM (MSSM), on which we now focus.

The mass matrix of the CP-even Higgs system of the MSSM contains three free parameters ---two masses and one mixing angle, or equivalently $m_A$, $\tan\beta$, and the well-known top--stop radiative correction $\Delta_t$. Identifying the 750 GeV resonance with the CP-even component of the heavier doublet, the masses, mixing, and couplings of all the Higgs states are determined as functions of $\tan\beta$ alone, which remains the only free parameter of the model.

The mixing angle between the two doublets, in the basis where one of the states takes all the vacuum expectation value, reads (see e.g. \cite{Barbieri:2013hxa})
\begin{equation}
\sin^2\delta = \frac{m_Z^2\cos^2 2\beta + \Delta_t^2 - m_h^2}{m_H^2 - m_h^2},
\end{equation}
where $m_H = 750$ GeV and $m_h = 125$ GeV are the masses of the two physical states, and the radiative correction $\Delta_t$ is determined as a function of the masses and $\tan\beta$. The mixing is largest, $\delta \simeq 0.3$, for $\tan\beta = 1$, which is close to the edge of future sensitivity of the high-luminosity LHC to modified Higgs couplings \cite{Snowmass}.

Neglecting loop effects due to new (supersymmetric) particles coupled to the Higgs bosons, the production cross section and branching ratios of $H$ are also determined. We have already shown in Sec.~\ref{sec:simplemodel} that with modified couplings to SM particles alone it is not possible to reproduce the diphoton excess. In this simple case, one finds the highest values of the $\gamma\gamma$ signal strength, at the level of only $\sim 10^{-2}$ fb, for very low values of $\tan\beta$, where also the production cross section is the largest.

Contributions from additional (supersymmetric) particles are required in order to further enhance the $\gamma\gamma$ rate. For low $\tan\beta$ the width of the 750 GeV state reaches 10--20 GeV, and Fig.~\ref{fig:largewidth} (bottom) shows that very large effective couplings to gluons and photons are needed in this case. Furthermore, since the branching fraction into $t\bar t$ is close to 1, also direct constraints from $t\bar t$ resonance searches are relevant, and require $c_G\lesssim 7$ and therefore $c_{\gamma\gamma}\gtrsim 40$.
On the other hand, the width of $H$ reaches its minimum of around a GeV for $\tan\beta \simeq 6$--$8$. In this case, as can be seen in Fig.~\ref{fig:largewidth} (top), smaller values of $c_G$ and $c_{\gamma\gamma}$ are needed.
However, loops of top squarks can increase the sizable only marginally \cite{djouadi}. $\mathcal{B}(H\to\gamma\gamma)$ can get a more significant contribution from light charginos \cite{djouadi}, but still at a level which is not enough to reproduce the observed signal strength.

It therefore looks difficult to accommodate the observed excess in the MSSM.
A generic two Higgs doublet model of type II, on the other hand, with the addition of new light charged and colored states coupled to the Higgs bosons, can easily accommodate a diphoton signal compatible with the excess. In this case one simply has to satisfy all the constraints as described in Sec.~\ref{wide}, and the $c_G$ and $c_{\gamma\gamma}$ coefficients can be estimated as in Eq.~\eqref{eq:matching}.

\section{Conclusion}

After analyzing the very first data at $13$~TeV collision energy, both ATLAS and CMS collaborations have recently reported a tantalizing excess in the diphoton invariant mass spectrum around $\sim 750$~GeV. When interpreting the excess as a scalar (or pseudo-scalar) resonance, produced dominantly via gluon-gluon fusion, no tension with $8$~TeV analyses is found. On the contrary, as shown in Fig.~\ref{fig:combination}, a slight excess observed at $8$~TeV in CMS nicely dovetail with the recent excess, contributing to the combined signal strength
\begin{equation}
\mu_{\rm{13 TeV}} = (4.6\pm 1.2) ~\rm{fb}~.
\end{equation}
We introduced, in Sec.~\ref{sec:simplemodel}, an effective parameterization of scalar and pseudo-scalar interactions with the SM fields, and computed the relevant production cross sections and decay widths in terms of the effective couplings. Assuming the production cross section to be dominated by gluon fusion, we showed that the top quark and $W$ contributions in the loop are not sufficient to explain the excess, requiring new colored and charged particles to exist.
In Sec.~\ref{sec:constraints}, we did a survey of all potentially relevant resonance searches for a neutral scalar at the LHC Run-I, summarizing the limits on $\sigma\times\mathcal{B}$ in Table~\ref{tab:constraints}. These, on the other hand, imply upper limits on the size of potential decay modes as shown in Fig.~\ref{fig:br-limits}.  Working in the effective framework, we identified, in general terms, two phenomenologically distinct scenarios based on the assumptions as regards the total decay width, to which we refer as ``loop-only'' and ``extra-width'' scenario.

In the ``loop-only'' scenario, the main assumption is that the new resonance mainly couples to the SM gauge bosons at loop level, and thus the total decay width is in the MeV range. The excess, in this case, can easily be explained for $\mathcal{O}(1)$ effective couplings $c_G$ and $c_{\gamma \gamma}$, while remaining in agreement with all other data. In Sec.~\ref{sec:models}, we give two specific examples that match to this scenario, namely: a model with a single vector-like quark generating the effective $g g S$ and $\gamma \gamma S$ couplings, and a setup in which the pseudo-scalar singlet, along with the Higgs, arises as a pseudo-Nambu--Goldstone boson of some spontaneous symmetry breaking, in which case the required couplings are generated by a combination of Wess--Zumino--Witten terms and loops of composite fermions.

The ``extra-width'' scenario assumes that there are additional tree-level decay channels that dominate over the loop induced ones. We investigate the phenomenological implications specifying $\Gamma_S=1$ or $20$~GeV. In the latter case, employing the limits from Table~\ref{tab:constraints}, we argue that the resonance can not dominantly decay to the SM gauge bosons, and identify the $t \bar{t}$ or monojet signatures at the LHC as a possible way out. This scenario can be realized in composite Higgs models (see Sec.~\ref{sec:comp}), where large $c_G$ and $c_{\gamma\gamma}$ couplings are obtained via interactions with composite resonances in the large representations (or large multiplicity) of the global symmetry of the strong sector.

In the intermediate regime ($\Gamma_S=1$~GeV), somewhat smaller photon and gluon couplings are required to fit the excess. However, also in this case strong constraints on the other couplings apply, making $t\bar t$ and invisible again the channels in which a sizable branching ratio is allowed. An example of a scenario of this kind could arise in a type-II two Higgs doublet model, like the MSSM. Here, knowing the mass of the heavier doublet leaves only one free parameter that determines the phenomenology at tree level. The total width of the heavier doublet is in the few GeV range for moderate values of $\tan\beta$, but larger widths can also be attained. Even in the optimal case of smallest total width, we estimate that loop corrections from supersymmetric particles are too small to explain the observed excess, calling for an interpretation beyond the MSSM.

Even though it is still too early to draw definite conclusions about the existence of a new resonance, our analysis shows that both experiments consistently point to a sizable excess at an invariant mass of around 750 GeV. Moreover, such a resonance could fit well in many reasonable scenarios beyond the SM. In all cases other light particles are required and interesting signatures are predicted to show up during the rest of Run-II at the LHC.

\section*{Acknowledgements}
 
We thank Gino Isidori and Riccardo Barbieri for useful discussions, and Michele Frigerio for comments on the draft. DB also thanks Diego Redigolo, Adam Falkowski, and Robert Ziegler for interesting discussions. This research was supported in part by the Swiss National Science Foundation (SNF) under contract 200021-159720.



\end{document}